\documentclass[10pt]{article}
\usepackage{amsmath, amssymb, latexsym, slashed, cancel, hyperref, accents, bm, xcolor}
\usepackage[numbers]{natbib}
\newcommand{\ie}{i.\,e.}

\newcommand{\lv}{Lorentz violation}
\newcommand{\bh}{black hole}
\newcommand{\sr}{superradiance}
\newcommand{\sn}{superresonance}

\newcommand{\eq}{equation}

\newcommand{\sptm}{spacetime}
\newcommand{\aco}{acoustic}

\newcommand{\dr}{dispersion relation}

\newcommand{\beq}{\begin{equation}}
\newcommand{\eeq}{\end{equation}} 
\newcommand{\ba}{\begin{align}}
\newcommand{\ea}{\end{align}}

\newcommand{\om}{\omega}

\newcommand{\Om}{\Omega}

\newcommand{\f}{\frac}
\newcommand{\tm}[1]{\textrm{#1}}
\newcommand{\ol}[1]{\overline{#1}}
\newcommand{\nn}{\nonumber}
\newcommand{\rt}{\rightarrow}

\title{Acoustic superradiance in a slightly viscous fluid}
\author{Oindrila Ganguly\footnote{lagubhai@gmail.com} \\
\textit{Institute of Physics, Bhubaneswar, Odisha 751005}}
\date{}
\begin{document}
\maketitle

\begin{abstract}
The acoustic analogue of the geometry of curved \sptm\ realised in a classical fluid becomes Lorentz violating in the presence of viscosity. 
We study how this effective  \lv\ affects acoustic \sr\ from the ergosphere of a rotating sonic \bh\ formed in such a fluid. It turns out that \lv\ imposes an upper bound on frequencies of acoustic perturbations that can get scattered from the ergosphere. Incidentally, this upper bound is same as that on the spectrum of superradiant frequencies. This study also reveals how \sr\ is in general modified when the wave propagates through a dispersive and dissipative medium. Our result is valid only upto linear order in the coefficient of viscosity and is thus a first approximation to the full solution, focussing on the key qualitative features. 
\end{abstract}

\section{Introduction} \label{sec:intro}

In any  wave scattering process, the total energy  carried by reflected and transmitted waves is less than or equal to the energy originally brought in by the incident wave.
However, it was discovered by Zeldovich \cite{zeldovich1971,zeldovich1972} that modes of scalar or electromagnetic radiation falling on a cylinder made of absorbing material which rotates about its axis of symmetry with a frequency $\Omega$ are actually amplified when
\begin{align}
\omega < m ~\Omega~. \label{eq:srcrit}
\end{align}
Here, $\omega$ is the frequency of the incident wave and $m$ is the projection of its angular momentum along the axis of rotation. This phenomenon is known as superradiance or rotational \sr\ (to distinguish it from Dicke \sr\ \cite{dicke1954} which refers to amplification of radiation due to coherence in the emitting medium) because the outgoing wave carries away more energy than was originally brought in. In fact, rotational \sr\ is a general feature of any macroscopic, rotating, dissipative system with internal degrees of freedom, causing transfer of energy from one medium to another usually  stimulated by wave scattering. Classically, the horizon of a black hole is a perfect absorber and it was predicted by Misner et al \cite{zeldovich1971,zeldovich1972,misner1972,starobinskii1973}  that low frequency waves scattered off the horizon of a rotating (Kerr) black hole would be similarly amplified at the cost of black hole mass and angular momentum. 
The condition for this rotational superradiance is the same as  \autoref{eq:srcrit}, with $\Omega$ replaced by $\Omega_H$, the angular velocity of the \bh\ horizon:
\begin{align}
\omega < m ~\Omega_{\textrm H}~. 
\label{eq:bhsr}
\end{align}
If quantum nature of the fields is further considered then \bh s can be shown to evaporate spontaneously even in the absence of rotation \cite{hawking1975}. This kind of spontaneous emission from black holes is known as Hawking radiation. Unfortunately, the signals of both kinds of radiation are very feeble and so, in spite of being theoretically predicted almost forty years ago, it has been impossible to confirm their presence observationally.

Interestingly, \sr\ and Hawking radiation can be completely attributed to the influence of \sptm\ geometry on the evolution of classical or quantum fields  without any need to refer to the dynamical equations of Einstein. This intrinsic feature permits us to construct alternate low energy models based on relatively more accessible systems like classical fluids, superfluids, Bose Einstein condensates etc. that can map  the geometry of a curved \sptm , albeit in certain limits. This field of study was pioneered by Unruh \cite{unruh1981} when he observed that  in an inviscid, barotropic, irrotational fluid flowing in flat 3-space, the \aco\ perturbations $\psi_a$ in the velocity potential satisfy an \eq\ identical to the massless Klein Gordon \eq\ in a curved background:
\begin{align}
\square \psi_a 
\equiv \frac{1}{\sqrt{- g}}
~\partial_{\mu}
\left(
\sqrt{- g}~
g^{\mu\nu}~\partial_{\nu}\right)~\psi_a
= 0~. 
\label{eq:akg}
\end{align}
Here, $\bm g$ is an effective Lorentzian  `\aco\ metric' that can be perceived only by the field $\psi_a$.  All through this article, a $b$ in the subscript denotes bulk or background fields while a subscript $a$ refers to linear acoustic perturbations over them. Its explicit form is,
\begin{align}
ds^2 
 \equiv g_{\mu\nu}~ dx^\mu~ dx^\nu 
 =
- ~ \frac{\rho_{\tm b}}{c_{\tm s}} 
\left[
\left( c_{\tm s}^2-v_{\tm b}^2 \right) dt^2 - \delta_{\tm{ij}}dx^{\tm i}~dx^{\tm j} + 2 (v_{\tm b})_{\tm i} dt~dx^{\tm i}
\right]~.
\label{eq:acint}
\end{align}
Note that the acoustic metric is distinct from the metric of the physical spacetime in which the fluid flows. 
Analogue models of gravity enjoy the advantage that the effective \sptm\ is completely determined by the bulk density $\rho_b$ and bulk velocity $\vec v_b$ of the fluid provided the fluid is barotropic and irrotational \cite{visser1998}. Here, $c_s$ denotes the local speed of sound defined by $c_s^2 = \partial p_b/\partial \rho_b$. By properly tuning these variables, it is possible to create acoustic analogues of various geometries including the \sptm\ outside a \bh .

A flow that can suitably mimic a rotating \bh\ must have a radially inward plus a tangential component of the velocity such that it is locally vorticity free \cite{visser1998}. These properties are inherent to the draining bathtub flow with a sink at the origin. Without loss of generality, we can work with a fluid flowing in two dimensions with a bulk velocity 
\begin{equation}
\vec v_{\tm b} = -\frac{A}{r}~\hat{r}+\frac{B}{r}~\hat{\phi}~. \label{eq:dbvel}
\end{equation}
$A, B$ are real, positive constants and $(r,\phi)$ are plane polar coordinates. $\vec v_{\tm b}$ is not only locally irrotational $(\vec\nabla \times \vec v_{\tm b} = 0)$ but is also locally divergence free $(\vec\nabla \cdot \vec v_{\tm b} = 0)$. 
The corresponding $(2+1)$ dimensional metric, referred to as a draining bathtub metric, is,
\begin{align}
{ds}_{\tm {DB}}^2
= - ~ \frac{\rho_{\tm b}}{c_{\tm s}}
\biggl[
\left(c_{\tm s}^2-\frac{A^2+B^2}{r^2}\right)
{dt}^2
- {dr}^2 
- r^2~{d\phi}^2  
- 
\frac{2~A}{r}~dr~dt
+ 2B~d\phi~dt
\biggr]~. 
\label{eq:dbmetric}
\end{align}
Clearly, the draining bathtub metric does not exactly correspond to a Kerr geometry, not even to a section of it. What is important is that, the two are \textit{qualitatively similar}.
This metric is also stationary and axisymmetric, thereby possessing symmetries corresponding to time translations generated by a timelike Killing vector field $\bm\partial_t$ and planar rotations generated by a spacelike Killing vector field $\bm\partial_\phi$.
On the 2-surface at $r_h = A / c_s$ in this flow, the fluid velocity is everywhere inward pointing and the radial component of the fluid velocity equals the local speed of sound. In the acoustic geometry, this surface, called a sonic or acoustic horizon,  behaves as an outer trapped surface and can be identified with the \textit{future event horizon} of the black hole. Thus, an \aco\ \bh\ or a dumb hole is formed. The radius of the ergosphere is determined by the vanishing of the metric component $g_{tt}$ which occurs at $r_e = \sqrt{A^{2}+B^{2}}/c_{\tm s}$. 
In  \cite{basak2003a,basak2003b}, a detailed derivation of the condition for  \sr\ of acoustic perturbations, also known as \sn ,
from a rotating dumb hole has been presented. The condition turns out to be the same as that of \autoref{eq:bhsr}.
%
%

This assumption of zero viscosity is necessary for the following two reasons:
\begin{enumerate}
	\item to ensure that viscous effects do not, locally, generate vorticity;
	\item to make the emergent \aco\ geometry Lorentzian.
\end{enumerate}
The latter is evident from the way viscous contributions alter the evolution of $\psi_a$:
\begin{align}
\square~\psi_a 
= - \frac{4}{3} 
\frac{\nu}{\rho_b c_s}
\left({\partial_t} + \vec v_b \cdot \vec \nabla\right)
\left(\nabla^2 \psi_a \right)~, \label{eq:kgnu}
\end{align}
$\nu$ denoting the coefficient of kinematic viscosity of the fluid. (For a pedagogical derivation of the acoustic metric in the inviscid and viscous case, please refer to \cite{visser1998}.) Here, the left hand side involves the general curved Lorentzian acoustic metric $\bm g$ while terms on the right hand side couple to flat metric of three space -- evidently violating Lorentz covariance of the equation. Moreover, new terms of $O(|\vec k|^3)$ and higher appear in the dispersion relation of the acoustic perturbations (as illustrated in \autoref{sec:acdr}), 
$\vec k$ being the three momentum of the perturbation $\psi_a$, also implying a breakdown of local acoustic Lorentz invariance. 

In the present article, we relax the very assumption of inviscidity in order to explore how \aco\ \sr\ is affected by Lorentz violation of the analogue \sptm . Physically, viscosity is known to have dissipative and dispersive effects on the propagation of sound waves through a fluid. So, it is expected also to modify the process of \aco\ \sr\  from a rotating dumb hole.  In \autoref{sec:sr}, we present an asymptotic solution to the problem in a limit where certain simplifying assumptions about the governing equations can be used. A condition same as \autoref{eq:bhsr} is obtained with the difference that no wave with frequency greater than $m~\Om_H$ is now allowed to undergo scattering from the ergosphere. 
We end this article with a discussion of the limitations of our result and planned improvements.

\section{Derivation of acoustic dispersion relation} \label{sec:acdr}

A dispersion relation expresses the momentum of a wave as a function of its energy. Let us assume $\psi_a$ to be a plane wave
\begin{align}
\psi_{\tm a} (\mathbf x) = A(\vec x) e^{i (\omega t - \vec k \cdot \vec x)}  \label{eq:eik}
\end{align}
obeying the eikonal approximation according to which the amplitude $A(\vec x)$ is taken to be a slowly varying function compared to the exponential. As a simplifying measure, we ignore derivatives of the metric. Substituting this into  \autoref{eq:kgnu} and using the explicit form of the metric, we obtain a quadratic \eq\ in $\left(\omega - \vec v_{\tm b} \cdot \vec k \right)$ \cite{visser1998}:
\begin{align*}
\left(\omega - \vec v_{\tm b} \cdot \vec k \right)^2 + i~\f{4\nu}{3} \left(\omega - \vec v_{\tm b} \cdot \vec k \right) |\vec k|^2
- c_{\tm s}^2 |\vec k|^2 = 0
\end{align*}
whose roots give the dispersion relation
\begin{align}
\omega =  
\vec v_{\tm b}\cdot \vec k \pm 
\sqrt{c_{\tm s}^2 |\vec k|^2 - \left({2\nu |\vec k|^2\over3}\right)^2 } 
- i {2\nu |\vec k|^2\over3}~. 
\label{eq:drnu}
\end{align}
In this non-linear dispersion relation, the $\nu$ dependent term under the big square root gives rise to dispersion while dissipative effects of viscosity are brought in by the imaginary term. A modified \dr\ is known to be a signature of \lv . However, the dispersive and dissipative acoustic Lorentz violating terms in \autoref{eq:drnu} contribute only at high momenta $\vec k$. This agrees intuitively with the fact that at high momenta the continuum fluid model breaks down thereby invalidating the assumption of a continuous `acoustic spacetime' \cite{visser1998}.


\section{Acoustic superradiance in a viscous fluid}\label{sec:sr}

Gravitational superradiance can occur  in the axisymmetric, stationary spacetime outside a rotating \bh\ because the ergosphere, \ie\ the region where the time translation Killing vector becomes spacelike, extends outside the event horizon to the domain of outer communication. As discussed in \autoref{sec:intro}, a draining bathtub flow provides an ideal arena for mapping such a geometry. 
To keep calculations simple, we restrict the coefficient of kinematic viscosity of the fluid to be sufficiently small ($\nu << 1$) so that terms of $O(\nu^2)$ and higher may be neglected with respect to the contribution at $O(\nu)$  and also assume that the background density $\rho_{\tm b}$ remains constant. This, in turn, implies the constancy of the bulk pressure (owing to barotropicity) and the local speed of sound. Thus, a further simplification becomes possible. We rescale the dimensions and set the local speed of sound to unity \ie\ $c_{\tm s} = 1$, for the rest of this article. Under these conditions, the metric of \autoref{eq:dbmetric} becomes,
\begin{align}
{ds}_{DB}^2
= - \rho_b
\biggl[
\left(1-\frac{A^2+B^2}{r^2}\right)
{dt}^2
-{dr}^2-r^2~{d\phi}^2  
-
\frac{2~A}{r}~dr~dt
+ 2B~d\phi~dt
\biggr] ~.
\label{eq:dbmetc1}
\end{align}
The viscous wave equation \eqref{eq:kgnu} when written out explicitly in the $(t, r, \phi)$ coordinates takes the form
\begin{align}
\biggl[
& \, 
- \partial_t^2 + 
\left(1 - \frac{A^2}{r^2}\right)
\partial_r^2
+ \frac{1}{r^2}
\left(1 - \frac{B^2}{r^2}\right)
\partial_\phi^2 
+ \frac{2A}{r}\partial_t\partial_r         
- \frac{2B}{r^2}\partial_t\partial_\phi   
+ \frac{2AB}{r^3}\partial_\phi\partial_r  
\left\{
\frac{1}{r}
\left(1 - \frac{A^2}{r^2}\right)
+ \frac{2A^2}{r^3}
\right\}
\partial_r 							\nonumber \\
& \,
- \frac{2AB}{r^4}\partial_\phi
\biggr]
\psi_a 						
= - \frac{4 \nu}{3}
\biggl[
- \frac{A}{r}\partial_r^3
+ \frac{B}{r^4}\partial_\phi^3 
- \frac{A}{r^3}\partial_r\partial_\phi^2
+ \frac{B}{r^2}\partial_r^2\partial_\phi   
+ \frac{1}{r^2}\partial_\phi^2 \partial_t  
+ \partial_r^2\partial_t                      
- \frac{A}{r^2}\partial_r^2
+ \frac{2A}{r^4}\partial_\phi^2 
+ \frac{B}{r^3}\partial_r\partial_\phi  		\nonumber \\
& \,
+ \frac{1}{r}\partial_r\partial_t           
+ \frac{A}{r^3}\partial_r
\biggr]
\psi_a ~.
\label{eq:eomnu}
\end{align}
Now, the massless Klein Gordon equation is known to allow complete separation of the variables in the Kerr metric \cite{carter1968,brill1972}. The same holds for the Klein Gordon equation with a mass term. As a result, the inviscid wave equation in a draining bathtub metric, obtained by setting $\nu = 0$ in \autoref{eq:eomnu} is also separable and the $(\phi, t)$ dependence is given by  the usual eigenfunctions appropriate to an axially symmetric and stationary background geometry, namely, $exp(-i\omega t + i m \phi)$ \cite{basak2003a,basak2003b}. Here, $\omega$ and $m$ are the frequency and orbital momentum of the wave  parallel to the symmetry axis.

In \autoref{eq:eomnu}, the highest order derivatives $\partial_r^3, \partial_{\phi}^3$ are suppressed by the parameter $\nu$ multiplied by a factor which is always less than one because we are interested in a solution outside the acoustic horizon, \ie , for $r \in (\textrm A, \infty)$. So, let us start by assuming a separable solution to the viscous wave \autoref{eq:eomnu}:
\begin{align*}
\psi_{\tm a}(t,r,\phi) = T(t)R(r)\Phi(\phi)~.
\end{align*}
We tune the flow parameters $A, B$ such that $B<<A$. This helps us go to the limit where the term $B \partial_{\phi}^3 \psi_{\tm a}$ in \autoref{eq:eomnu} may be ignored in comparison to $2A \partial_{\phi}^2  \psi_{\tm a}$. As a result, the highest order of axial derivative occuring in  \autoref{eq:eomnu} is reduced to two. We employ the ans\"atz:
\begin{align}
\psi_{\tm a}(t,r,\phi) = R(r) e^{-i \omega t + i m \phi} ~.  \label{eq:an}
\end{align}
Following the standard steps, we can find the separated radial \eq\ to be
\begin{align}
& \,
- \f{4\nu}{3} \f{A}{r} \partial_r^3
+ \biggl[ 
1 - \frac{A^2}{r^2} 
- \f{4\nu}{3}
\biggl\{
\frac{A}{r^2}
+ i \biggl( \omega - \f{Bm}{r^2} \biggr)
\biggr\}
\biggr]
\f{d^2}{dr^2} R(r)          \nonumber \\ 
& \,
+
 \biggl[
\frac{1}{r}
\biggl(1 - \frac{A^2}{r^2}\biggr)
+ \frac{2A^2}{r^3}
- \frac{i 2A}{r}
\biggl( \omega - \frac{Bm}{r^2}\biggr)          
+ \f{4\nu}{3}
\biggl\{
\frac{A}{r^3}
+ \frac{A m^2}{r^3}
- \frac{i}{r}
\biggl(\omega - \frac{B m}{r^2}\biggr) 
\biggr\}
\biggr] 
\f{d}{dr}R(r)   			\nonumber \\ 
& \,
+
\biggl[
\biggl( \omega - \frac{Bm}{r^2}\biggr)^2
-\f{m^2}{r^2}
- \frac{i 2 ABm}{r^4}          
- \f{4\nu}{3}
\biggl( \frac{2Am^2}{r^4} - \frac{i m^2 \omega}{r^2} \biggr)
\biggr] 
R(r) = 0 ~.
\label{eq:3r}
\end{align}
This is a linear third order ordinary homogeneous differential equation in $R(r)$. Ideally, it should be possible to analytically find a general solution to this linear \eq , but the problem becomes rather intractable.  
So, to get a first approximation to the complete answer, we instead study the asymptotic form of \autoref{eq:3r} in the limit where 
\begin{align*}
d_r^3 R(r) << \f{1}{r} d_r^2 R(r)~.
\end{align*} 
This is equivalent to requiring 
\begin{align*}
\partial_r^3 \psi_{\tm a} << \f{1}{r} \partial_r^2 \psi_{\tm a}~.
\end{align*} 
Thus, restricting the parameter space by the criterion $B<<A$, employing ans\"atz  of \autoref{eq:an} and finally constraining $\psi_{\tm a}$ to vary slowly with respect to $r$, we are able to reduce \autoref{eq:3r} to the following second order, linear differential equation:
\begin{align}
 & \, 
\biggl[ 
 1 - \frac{A^2}{r^2} 
- \f{4\nu}{3}
\biggl\{
\frac{A}{r^2}
+ i \biggl( \omega - \f{Bm}{r^2} \biggr)
\biggr\}
\biggr]
\f{d^2}{dr^2} R(r) 				        \nonumber \\
& \,
+  
\biggl[
\frac{1}{r}
\biggl(1 - \frac{A^2}{r^2}\biggr)
+ \frac{2A^2}{r^3}
- \frac{i 2A}{r}
\biggl( \omega - \frac{Bm}{r^2}\biggr)          
+ \f{4\nu}{3}
\biggl\{
\frac{A}{r^3}
+ \frac{A m^2}{r^3}
- \frac{i}{r}
\biggl(\omega - \frac{B m}{r^2}\biggr) 
\biggr\}
\biggr] 
\f{d}{dr}R(r)   							\nonumber \\
& \,
+
\biggl[
\biggl( \omega - \frac{Bm}{r^2}\biggr)^2
-\f{m^2}{r^2}
- \frac{i 2 ABm}{r^4}          					
- \f{4\nu}{3}
\biggl( \frac{2Am^2}{r^4} - \frac{i m^2 \omega}{r^2} \biggr)
\biggr] 
R(r) = 0~. 
\label{eq:2r}
\end{align}
But this \eq\ can be lent a simpler form by defining a new radial coordinate $r_*$ called the tortoise coordinate as
\begin{align}
\biggl[ 
1 - \frac{A^2}{r^2} 
- \f{4\nu}{3}
\biggl\{
\frac{A}{r^2}
+ i \biggl( \omega - \f{Bm}{r^2} \biggr)
\biggr\}
\biggr]\f{d}{dr}
\equiv \f{d}{dr_*}   \label{eq:rstar}
\end{align}
and a new function $\ol R(r_*)$ as $\ol R(r_*) = R(r)$.
\footnote{When $\nu = 0$, $r_*$ remains real.}
This definition makes the tortoise coordinate a complex variable and the differential equation is now defined on the complex $r_*$ plane instead of the real $r$ axis. The bar over $R$ in $\ol R$ is used to denote its dependence on $r_*$ and in the following formulae we shall place a bar over all other functions of $r_*$. Under the transformation $r\rightarrow r_*$ and $R(r) \rightarrow \ol R(r_*)$, \autoref{eq:2r} becomes,
\begin{align}
& \,
\f{d^2}{dr_*^2}\ol R(r_*)					          
+ \biggl[
\frac{1}{r}
\biggl(1 - \frac{A^2}{r^2}\biggr)
- \frac{i 2A}{r}
\biggl( \omega - \frac{Bm}{r^2}\biggr)  
+ \f{4\nu}{3}
\biggl\{
\frac{A m^2}{r^3}
- \frac{A}{r^3}
- \frac{i}{r}
\biggl(\omega - \frac{B m}{r^2}\biggr) 
+\f{i 2Bm}{r^3}
\biggr\}
\biggr] \f{d}{dr_*}  \ol R(r_*)   \nonumber \\
& \,
+
\biggl[ 
1 - \frac{A^2}{r^2} 
- \f{4\nu}{3}
\biggl\{
\frac{A}{r^2}
+ i \biggl( \omega - \f{Bm}{r^2} \biggr)
\biggr\}
\biggr]  								          
\biggl[
\biggl( \omega - \frac{Bm}{r^2}\biggr)^2 	
- \f{m^2}{r^2}								
- \frac{i 2 ABm}{r^4} 
- \f{4\nu}{3}
\biggl( \frac{2Am^2}{r^4} - \frac{i m^2 \omega}{r^2} \biggr)
\biggr] 											
\ol R(r_*)                 \nn \\
& \,
= 0 ~.
\label{eq:2rstar}
\end{align}
Note that the coefficients here remain functions of $r$. They have not been changed because we do not know explicitly $r(r_*)$. In fact, we do not need to know this inverse transformation. If we denote the coefficients of $d\ol R(r_*)/ dr_*$ and $ \ol R(r_*)$ by $a(r) = \ol a(r_*)$ and $b(r) = \ol b(r_*)$ respectively, then \autoref{eq:2rstar} is simply of the form
\begin{align}
\f{d^2}{dr_*^2}\ol R(r_*)
+ a(r) \f{d}{dr_*}\ol R(r_*) 
+ b(r)\ol R(r_*) = 0 ~.             \label{eq:2small}
\end{align}
We wish to write this as a one dimensional equation with an effective potential by defining another radial function $\ol\zeta(r_*)$. In other words, we plan to make a transformation from $\ol R(r_*) \rightarrow \ol\zeta(r_*)$ in a way that will make the coefficient of $d\ol\zeta/dr_*$ zero. If
\begin{align*}
\ol R(r_*) \equiv \tm{exp}\biggl[- ~ \f{1}{2} \int^{r_*} ds_* \ol a(s_*) \biggr] \ol\zeta(r_*)
\end{align*}
then \autoref{eq:2small} is indeed reduced to 
\begin{align}
\f{d^2}{dr_*^2}\ol\zeta
+ \ol k^2(r_*)\ol\zeta = 0				\label{eq:2schr} 
\end{align}
where retaining terms upto $O(\nu)$,
\begin{align}
\ol k^2(r_*) 
= k^2(r)										
& \,
\approx 
\biggl( \omega - \frac{Bm}{r^2}\biggr)^2
+
\biggl(1 - \f{A^2}{r^2} \biggr)
\biggl\{
\biggl(1 - \f{A^2}{r^2} \biggr) \f{1}{4r^2}		
-\f{m^2}{r^2} -\f{A^2}{r^4}
\bigg\}										\nonumber\\
& \,
- \f{2\nu}{3}
\biggl[
\biggl(1 - \f{A^2}{r^2} \biggr)
\biggl\{
\f{3A}{r^4} + \f{2Am^2}{r^4}
+  \frac{i}{r^2}
\biggl(\omega - \frac{B m}{r^2}\biggr) 		
-  \f{i 2 m^2 \omega}{r^2} - \f{i 6Bm}{r^4}
\biggr\}								\nonumber \\
& \,
+ 2i 
\biggl(\omega - \frac{B m}{r^2}\biggr)^3
+ \f{2A}{r^2}
\biggl(\omega - \frac{B m}{r^2}\biggr)^2		\nonumber \\
& \,
+ \biggl(\omega - \frac{B m}{r^2}\biggr)
\biggl\{
\f{4ABm}{r^4} - \f{i2m^2}{r^2}
- \f{i2A^2}{r^4}(1+m^2)
\biggr\}								
- \f{2Am^2}{r^4} - \f{2A^3}{r^6}	
\biggr]								\label{eq:2k2}				
\end{align}
Any differential equation of the form of \autoref{eq:2schr} (with vanishing coefficient of the first derivative) is known to have a constant Wronskian. This allows us to solve the asymptotic forms of \autoref{eq:2schr} in the limiting cases of $r\rightarrow A$ (near the acoustic horizon) and  $r\rightarrow \infty$ and equate the corresponding Wronskians in order to deduce the conditions for superresonance (following \cite{basak2003a,basak2003b}).

\subsection*{Near horizon}
The surface at $r=A$ acts as an acoustic horizon rotating at the angular velocity $\Omega_{\tm H} = \f{B}{A^2}$.
Owing to the restrictions on the parameter space $(B<<A)$, the angular velocity of the horizon is quite small. Now, as the radial coordinate $r\rightarrow A$, let the functions $\ol\zeta(r_*)\rightarrow\ol\zeta_{\tm{hor}}(r_*)$ and
\begin{align}
k^2(r) \rt k_{\tm{hor}}^2(r) 		
\approx 
(\omega - m \Omega_{\tm H})^2		
- \f{4\nu}{3}
& \,
\biggl[
(\om - m \Om_{\tm H})^2\f{1}{A}
+ (\om - m \Om_{\tm H})\f{2m\Om_{\tm H}}{A}
- \f{1+m^2}{A^3}  &					\nn\\
& \, 
+ i (\om - m \Om_{\tm H})^3
-i (\om - m \Om_{\tm H})\f{1+2m^2}{A^2}
\biggr]	&
\label{eq:2k2h}
\end{align}
from \autoref{eq:2k2}. Therefore, 
\begin{align}
 k_{\tm{hor}} \approx 
 (\omega - m \Omega_{\tm H})		
- \f{2\nu}{3A}
\biggl\{
(     \om + m \Om_{\tm H}     )
- \f{1+m^2}{A^2 (\omega - m \Omega_{\tm H})}
\biggr\}					
+ i \f{2\nu}{3} 
\biggl\{
\f{1+2m^2}{A^2}
- (\om - m \Om_{\tm H})^2
\biggr\}~.
\label{eq:2kh}
\end{align}
A solution to the approximate near horizon differential equation
\begin{align}
\f{d^2}{dr_*^2}\ol\zeta_{\tm{hor}}
+ \ol k_{\tm{hor}}^2\ol\zeta_{\tm{hor}} = 0				\label{eq:2schrh} 
\end{align}
can easily be written down as
\begin{align}
\ol\zeta_{\tm{hor}1}(r_*) = T_{\om \tm m} e^{-i k_{\tm{hor}} r_*}~.  \label{eq:2schrhsol1}
\end{align}
Here, $T_{\om \tm m}$ is the transmission coefficient and we have incorporated the boundary condition that the group velocity of the wave for $r\rt A$ is directed towards the trapping surface. Another linearly independent solution to \autoref{eq:2schrh} can be
\begin{align}
\ol\zeta_{\tm{hor}2}(r_*) = T_{\om \tm m}^* e^{i k_{\tm{hor}} r_*}~.	\label{eq:2schrhsol2}
\end{align}
The corresponding Wronskian $W_{\tm{hor}}$ is given by,
\begin{align}
W_{\tm{hor}} 
= \ol\zeta_{\tm{hor}1}\f{d}{dr_*}\ol\zeta_{\tm{hor}2} - \ol\zeta_{\tm{hor}2}\f{d}{dr_*}\ol\zeta_{\tm{hor}1}		
= 2 i k_{\tm{hor}} |T_{\om \tm m}|^2 ~.				\label{eq:2whor}
\end{align}
%

\subsection*{Near infinity}
In the asymptotic region where  $r\rt \infty$,  the function $\ol\zeta(r_*)\rightarrow\ol\zeta_{\tm{inf}}(r_*)$ and
\begin{align}
k^2(r) \rt k_{\tm{inf}}^2(r) 		
\approx 
\om^2 - i\f{4\nu}{3} \om^3
\approx 
\biggl(\om - i\f{2\nu}{3} \om^2 \biggr)^2		\label{eq:2k2in}
\end{align}
as before keeping only upto linear terms in $\nu$. So, 
\begin{align}
 k_{\tm{inf}} \approx \om - i\f{2\nu}{3} \om^2 ~.			
\label{eq:2kin}
\end{align}
A solution to the approximate differential equation
\begin{align}
\f{d^2}{dr_*^2}\ol\zeta_{\tm{inf}}
+ \ol k_{\tm{hor}}^2\ol\zeta_{\tm{inf}} = 0				\label{eq:2schrin} 
\end{align}
is
\begin{align}
\ol\zeta_{\tm{inf}1}(r_*) = R_{\om \tm m} e^{i k_{\tm{inf}} r_*} +  e^{- i k_{\tm{inf}} r_*}~.	\label{eq:2schrinsol1}
\end{align}
Here, $R_{\om \tm m}$ is the reflection coefficient in the sense of potential scattering and the amplitude of the incident wave is normalised to unity. Another linearly independent solution to \autoref{eq:2schrin} is
\begin{align}
\ol\zeta_{\tm{inf}2}(r_*) = R_{\om \tm m}^* e^{- i k_{\tm{inf}} r_*} + e^{i k_{\tm{inf}} r_*}~.			\label{eq:2schrinsol2}
\end{align}
So, the Wronskian
\begin{align}
W_{\tm{inf}} 
= \ol\zeta_{\tm{inf}1}\f{d}{dr_*}\ol\zeta_{\tm{inf}2} - \ol\zeta_{\tm{inf}2}\f{d}{dr_*}\ol\zeta_{\tm{inf}1}		
= 2 i k_{\tm{inf}} (1 - |R_{\om \tm m}|^2) ~.				
\label{eq:2win}
\end{align}
%

\subsection*{Comparison of Wronskians}
The Wronskian of \autoref{eq:2k2} being constant, we have
\begin{align}
\f{1 -  |R_{\om \tm m}|^2}{|T_{\om \tm m}|^2} & \ = \f{k_{\tm{hor}}}{k_{\tm{inf}}}~.				\label{eq:ww}
\end{align}
We need to determine the asymptotic form of the ratio $k_{\tm{hor}}/k_{\tm{inf}}$ when  $\nu << 1$. Using \autoref{eq:2kh} and \autoref{eq:2kin} and neglecting $O(\nu^2)$ or higher contributions, we arrive at
\begin{align}
 \f{k_{\tm{hor}}}{k_{\tm{inf}}} 		
& \,
\approx 
\f{1}{\om}
\biggl[
\omega - m \Omega_{\tm H}		
- \f{2\nu}{3A}
\biggl\{
\om + m \Om_{\tm H}
- \f{1+m^2}{A^2 (\omega - m \Omega_{\tm H})}
\biggr\}
\biggr]									\nn \\
& \,
+ i \f{2\nu}{3 \om} 
\biggl\{
\f{1+2m^2}{A^2}
- (\om - m \Om_{\tm H})^2
+ (\om - m \Om_{\tm H})\om
\biggr\}	~.				
\label{eq:kk}
\end{align}
\footnote{In an inviscid fluid, no such condition appears and scattering can occur for the entire range of frequencies. But there will be \sn\ only when $0 < \om < m \Om_\tm H$.}
However, from \autoref{eq:ww} it is clear that $k_{\tm{hor}}/k_{\tm{inf}}$ must be real. So,
\begin{align}
m \Om_{\tm H} (\om - m \Om_{\tm H}) = -  ~ \f{1+2m^2}{A^2}~.			\label{eq:real}
\end{align}
As the expression to the right of the equality is always negative, this equation puts the following constraint  on $\om$:
\begin{align*}
\om < m \Om_\tm H~.
\end{align*}
This must hold to \textit{allow scattering} of acoustic perturbations off the ergosphere of a dumb hole in a fluid with a small kinematic viscosity $\nu$ if the conditions specified above are true.

Now, when superresonance occurs, intensity of the reflected wave is greater than that of the incident wave \ie\ $|R_{\om \tm m}|^2 > 1$. \autoref{eq:ww} implies that a necessary condition for superresonance is $\f{k_{\tm{hor}}}{k_{\tm{inf}}} < 0	$, that is,
\begin{align}
(\omega - m \Omega_{\tm H})^2		
- \f{2\nu}{3A}
\biggl\{
(\om + 2m \Om_{\tm H})(\omega - m \Omega_{\tm H})
+ \f{m^2}{A^2}
\biggr\}
 > 0~.					
 \label{eq:ineq}
\end{align}
This is obtained by simplifying \autoref{eq:kk} with the help of \autoref{eq:real}. The roots of the corresponding equation
are given by Sridhar Acharya rule to be,
\begin{align*}
\om
\approx
\biggl(
1 +  \f{\nu}{A}
\biggr)  m \Omega_{\tm H}  		
\pm
\f{m}{A}
\sqrt{ 
 \f{2\nu}{3A}
}
\end{align*}
as $O(\nu^2)$ and higher contributions are ignored. Inserting this into the inequality of \autoref{eq:ineq}, we have
\begin{align*}
\bigg[
\om 
- \biggl(
1 +  \f{\nu}{A}
\biggr)  m \Omega_{\tm H}  		
- \f{m}{A}
\sqrt{ 
 \f{2\nu}{3A}
}\bigg]
\bigg[
\om 
- \biggl(
1 +  \f{\nu}{A}
\biggr)  m \Omega_{\tm H}  		
+ \f{m}{A}
\sqrt{ 
 \f{2\nu}{3A}
}\bigg] > 0 ~. 
\end{align*}
Since, $\om < m \Omega_{\tm H}$, the expression within the first set of brackets is negative for all values of $\om$. As a result, we must have
\begin{align}
\om 
< 
\biggl(
1 +  \f{\nu}{A}
\biggr)  m \Omega_{\tm H}  		
- \f{m}{A}
\sqrt{ 
 \f{2\nu}{3A}
}~.					\label{eq:om}
\end{align}
We must convince ourselves that this result is consistent with \autoref{eq:real} ( $\om < m \Omega_{\tm H}$). It will be so if
\begin{align*}
\f{m}{A}
\sqrt{ 
 \f{2\nu}{3A}} 
 &>
\f{\nu}{A} m \Omega_{\tm H}~, \\
 \f{2\nu}{3A}
&> \f{\nu^2 B^2}{A^4}~.
\end{align*}
The last step is obtained after dividing by the common non-zero factor $m/A$ and taking the square of both sides. We have also substituted the formula for 
$\Omega_{\tm H}$. This leads to the relation
\begin{align*}
\f{2}{3} > \f{\nu B^2}{A^3}
\end{align*}
which obviously is true because we have earlier made the choice that $\nu << 1$ and $B/A << 1$. Hence, we can conclude that in the limit where our assumptions hold good, sound waves will be scattered off the ergosphere of a rotating dumb hole when  $\om < m \Omega_{\tm H}$ and there will be superresonance for \textit{all such waves}.

\section{Summary and discussion}
Our primary motivation in pursuing this study has been to understand how the absence of Lorentz covariance of the acoustic wave \eq\  in emergent \aco\ \sptm , arising due to the consideration of viscosity of the fluid, influences \sn\ from a rotating dumb hole. Instead of tackling the full problem with all its complexity, we have made a number of simplifying assumptions to gain an estimate of these effects to a first order in $\nu$. The assumptions that we have employed are:
\begin{enumerate}
	\item Kinematic viscosity $\nu$ has been taken to be small enough 
	so that terms higher than $O(\nu)$ can be neglected;
	\item Bulk fluid flow is such that $B << A$ (tangential fluid velocity is negligible compared to the radial component) implying $\f{B}{r^4} \partial_\phi^3 \psi_{\tm a} << \f{2A}{r^4} \partial_\phi^2 \psi_{\tm a}~, r \in (A, \infty)$;
	\item $\psi_{\tm a} = R(r) exp [-i \om t + i m \phi], \om, m > 0 $ and $m$ takes integral values;
	\item Only regions of slow variation of $R(r)$ are considered. So, \autoref{eq:eomnu} reduces approximately to \autoref{eq:2r}. Here, $d_r^3 R$ has been neglected in comparison to $(1/r) d_r^2 R$.
\end{enumerate}
Provided these assumptions hold good, our results indicate that \aco\ \lv\ restricts the range of allowed frequencies $\om$ for a scattering process to be always less than $m \Omega_{\tm H}$. But, this exactly coincides with the range of frequencies that can undergo superradiance. So, all such solutions to \autoref{eq:eomnu} get amplified upon scattering by extracting rotational energy from the dumb hole. The assumptions made by us have somehow conspired to always suppress the frequency $\om$ below $m \Omega_{\tm H}$. It is instructive to explore how this constraint gets modified when a more general analysis of \autoref{eq:eomnu} is affordable.

Apart from being a study into Lorentz violation, it is by itself interesting to know the effects of viscosity on \sn . But the interest is not only theoretical. Owing to the requirement of zero viscosity, real fluids cannot be used to build sonic \bh\ models in the laboratory. One is forced to look into ideal fluids like superfluid helium. However, with an estimate of how viscosity modifies the outcome in a real fluid, it may become practically possible to deduct viscous effects appearing in real fluids and extrapolate a result to the ideal, inviscid case. 
This is also true of black hole analogues built using gravity waves in water, another promising candidate for simulating the spacetime outside a \bh . Long wavelength gravity waves in a shallow basin filled with a flowing fluid are also governed by the same \eq\ as a scalar field in curved \sptm\ \cite{schutzhold2002}. In fact, very recently, amplification of gravity waves after being scattered from a draining vortex has been observed in the laboratory \cite{torres2016}. However, the experimenters have not been able to confirm whether this amplification is due to a rotating analogue \bh\ or due to dissipation inside the vortex core, as discussed in \cite{cardoso2016}. Here, too, zero viscosity is a prior assumption in the  theoretical framework. But it holds only  approximately  when water is used in the experiments. So, it is instructive and, in fact, necessary to determine the impact of  viscosity on  \sr\ in this scenario. We plan to address this problem in  future.

\section{Acknowledgements}
The author is grateful to Parthasarathi Majumdar for his insightful guidance at every step of this problem and also thanks Nirupam Dutta for numerous discussions on this topic. 

\bibliographystyle{unsrtnat}
\bibliography{Bibliography}{}

\end{document}